\documentclass[prl, nofootinbib, aps, twocolumn, superscriptaddress,
showpacs, preprintnumbers, amsmath, amssymb, floatfix]{revtex4}

\usepackage{graphicx}

\newcommand{\Order}{{\cal O}}   % e.g. terms up to $\Order(g^2)$
\newcommand{\MeV}{\mathrm{MeV}}
\newcommand{\GeV}{\mathrm{GeV}}
\newcommand{\TeV}{\mathrm{TeV}}
\newcommand{\Mpc}{\mathrm{Mpc}}
\newcommand{\km}{\mathrm{km}}
\newcommand{\seconds}{\mathrm{s}}

\newcommand{\MPl}{\mathrm{M}_{\mathrm{P}}}
\newcommand{\gravitino}{{\widetilde{G}}}
\newcommand{\stau}{{\widetilde{\tau}_1}}

\newcommand{\scalartop}{{\widetilde{t}_1}}
\newcommand{\st}{{\tilde{\tau}_1}}

\newcommand{\neutralino}{{\widetilde \chi}^{0}_{1}}
\newcommand{\chargino}{{\widetilde{\chi}_1^{\pm}}}
\newcommand{\Higgs}{\mathrm{H}}
\newcommand{\mgr}{m_{\widetilde{G}}}

\newcommand{\ZB}{\mathrm{Z}}

\newcommand{\NLSP}{\mathrm{NLSP}}
\newcommand{\NTP}{\mathrm{NTP}}
\newcommand{\TP}{\mathrm{TP}}

\newcommand{\equil}{\mathrm{eq}}
\newcommand{\freezeout}{\mathrm{f}}
\newcommand{\CDM}{\mathrm{dm}}

\newcommand{\rad}{\mathrm{rad}}
\newcommand{\GUT}{\mathrm{GUT}}
\newcommand{\Reheating}{\mathrm{R}}
\newcommand{\TR}{T_{\Reheating}}
\newcommand{\TL}{T_{\mathrm{low}}}
\newcommand{\Color}{\mathrm{c}}
\newcommand{\Weak}{\mathrm{L}}
\newcommand{\Hypercharge}{\mathrm{Y}}

\begin{document}
% 
% ___ Preprint Numbers ___________________________________________________
%
\preprint{hep-ph/0612291}
\preprint{MPP-2006-141}
%
%
% ___ Preamble ______________________________________________________
%
\title{Constraints on the Reheating Temperature in Gravitino Dark
  Matter Scenarios}
\author{Josef~Pradler}
\email{jpradler@mppmu.mpg.de}
\affiliation{Max-Planck-Institut f\"ur Physik, 
F\"ohringer Ring 6,
D--80805 Munich, Germany}
\author{Frank Daniel Steffen}
\email{steffen@mppmu.mpg.de}
\affiliation{Max-Planck-Institut f\"ur Physik, 
F\"ohringer Ring 6,
D--80805 Munich, Germany}
%
%\date{\today}
%
% ___ Abstract _________________________________________________________
%
\begin{abstract}
  Considering gravitino dark matter scenarios, we study constraints on
  the reheating temperature of inflation.
  We present the gauge-invariant result for the thermally produced
  gravitino yield to leading order in the Standard Model gauge
  couplings.
  Within the framework of the constrained minimal supersymmetric
  Standard Model (CMSSM),
  we find a maximum reheating temperature of about $10^7~\GeV$ taking
  into account bound-state effects on the primordial $^6$Li abundance.
  We show that late-time entropy production can relax this constraint
  significantly.
  Only with a substantial entropy release after the decoupling of the
  lightest Standard Model superpartner, thermal leptogenesis remains a
  viable explanation of the cosmic baryon asymmetry within the CMSSM.
\end{abstract}
%
% 04.65.+e      Supergravity (see also 12.60.Jv Supersymmetric models)
%% 12.60.Jv     Supersymmetric models (see also 04.65.+e Supergravity)
% 14.80.Ly      Supersymmetric partners of known particles
%% 95.30.Cq     Elementary particle processes
%% 95.35.+d     Dark matter
%% 98.80.Cq     Particle-theory and field-theory models of the early Universe
%
%\keywords{}
\maketitle
\section{Introduction}

The observed flatness, isotropy, and homogeneity of the Universe
suggest that its earliest moments were governed by
inflation~\cite{Linde:1990,Kolb:vq}. The inflationary expansion is
followed by a phase in which the Universe is reheated.  The reheating
process repopulates the Universe and provides the initial conditions
for the subsequent radiation-dominated epoch.  We refer to the
reheating temperature $T_{\Reheating}$ as the initial temperature of
this early radiation-dominated epoch of our Universe.

The value of $T_{\Reheating}$ is an important prediction of inflation
models. While we do not have evidence for temperatures of the Universe
higher than $\Order(1~\MeV)$ (i.e., the temperature required by
primordial nucleosynthesis), inflation models can point to
$T_{\Reheating}$ well above
$10^{10}~\GeV$~\cite{Kolb:vq,Linde:1991km}.

In this Letter we consider supersymmetric (SUSY) extensions of the
Standard Model in which the gravitino $\gravitino$ is the lightest
supersymmetric particle (LSP) and stable because of R-parity
conservation. The gravitino LSP is a well-motivated dark matter
candidate.  As the gauge field of local SUSY transformations and the
spin-3/2 superpartner of the graviton, the gravitino is an unavoidable
implication of SUSY theories including gravity~\cite{Wess:1992cp}. Its
interactions are suppressed by inverse powers of the (reduced) Planck
scale $\MPl=2.4\times 10^{18}\,\GeV$. Its mass $\mgr$ results from
spontaneous SUSY breaking and can range from the eV scale up to scales
beyond the TeV region~\cite{Martin:1997ns}.

While any initial population of gravitinos must be diluted away by the
exponential expansion during inflation~\cite{Khlopov:1984pf},
gravitinos are regenerated in scattering processes of particles that
are in thermal equilibrium with the hot primordial plasma. The
efficiency of this thermal production of gravitinos during the
radiation-dominated epoch is sensitive to
$T_{\Reheating}$~\cite{Ellis:1984eq,Moroi:1993mb,Bolz:1998ek,Bolz:2000fu,Pradler:2006qh}.
Since the resulting gravitino density $\Omega_{\gravitino}^{\TP}$ is
bounded from above by the dark matter density $\Omega_{\CDM}$, upper
bounds on $T_{\Reheating}$ can be
derived~\cite{Moroi:1993mb,Asaka:2000zh,Roszkowski:2004jd,Cerdeno:2005eu,Steffen:2006hw}.
These bounds can be compared with predictions of the reheating
temperature $T_{\Reheating}$ from inflation models. Moreover,
$T_{\Reheating}$ is important for our understanding of the cosmic
baryon asymmetry.  For example, successful standard thermal
leptogenesis~\cite{Fukugita:1986hr} requires $T_{\Reheating}\gtrsim
10^9~\GeV$~\cite{Buchmuller:2004nz}.

We update the $T_{\Reheating}$ limits using the full gauge-invariant
result for the relic density of thermally produced gravitinos,
$\Omega_{\gravitino}^{\TP}$, to leading order in the Standard Model
gauge couplings~\cite{Pradler:2006qh}.
This allows us to illustrate the dependence of the bounds on the
gaugino-mass relation at the scale of grand unification $M_{\GUT}
\simeq 2 \times 10^{16}\,\GeV$.

We consider gravitino dark matter scenarios also in the framework of
the constrained minimal supersymmetric Standard Model (CMSSM) in which
the gaugino masses, the scalar masses, and the trilinear scalar
interactions are assumed to take on the respective universal values
$m_{1/2}$, $m_0$, and $A_0$ at $M_{\GUT}$.
Taking into account gravitinos from thermal production and from late
decays of the lightest Standard Model superpartner, we provide new
upper bounds on the reheating temperature in the $(m_{1/2},m_0)$ plane
for various values of $\mgr$.
Previous studies of $T_{\Reheating}$ constraints within the CMSSM used
the result of~\cite{Bolz:2000fu} to explore the viability of
$T_{\Reheating}\gtrsim 10^9\,\GeV$~\cite{Roszkowski:2004jd,Cerdeno:2005eu}.
Our study presents also scans for $T_{\Reheating}$ as low as
$10^7\,\GeV$ based on the result of~\cite{Pradler:2006qh} which
includes electroweak contributions to thermal gravitino
production~\cite{Pradler:inPreparation}.

In the considered CMSSM scenarios with the gravitino LSP, the
next-to-lightest SUSY particle (NLSP) is either the lightest
neutralino $\neutralino$ or the lighter stau $\stau$.\footnote{For
  simplicity, we consider $A_0=0$ in this work. For sizable $|A_0|$,
  also the lighter stop $\scalartop$ can be the
  NLSP~\cite{Boehm:1999bj,Ellis:2001nx}.}  Because of the extremely
weak interactions of the gravitino, the NLSP typically has a long
lifetime before it decays into the gravitino.  If these decays occur
during or after big-bang nucleosynthesis (BBN), the Standard Model
particles emitted in addition to the gravitino can affect the
abundance of the primordial light elements. Indeed, these BBN
constraints disfavor the $\neutralino$ NLSP for $\mgr\gtrsim
100~\MeV$~\cite{Feng:2004mt,Roszkowski:2004jd,Cerdeno:2005eu}. For the
slepton NLSP case, the BBN constraints associated with
hadronic/electromagnetic energy injection have also been estimated and
found to be much weaker but still significant in much of the parameter
space~\cite{Feng:2004mt,Roszkowski:2004jd,Cerdeno:2005eu,Steffen:2006hw}.

Only recently, it has been stressed that bound--state formation of
long-lived negatively charged particles with the primordial nuclei can
affect
BBN~\cite{Pospelov:2006sc,Kohri:2006cn,Kaplinghat:2006qr,Cyburt:2006uv}.
With the charged long-lived stau NLSP, these bound--state effects also
apply to the considered gravitino dark matter scenarios.  In
particular, a significant enhancement of $^6$Li production has been
found to imply severe upper limits on the $\stau$ NLSP abundance prior
to decay~\cite{Pospelov:2006sc}
which strongly restricts the mass spectrum in the $\stau$ NLSP
case~\cite{Steffen:2006wx}.
For generic parameter regions of the CMSSM, we show that this
constraint disfavors $T_{\Reheating} > 10^7~\GeV$ and thereby
successful thermal leptogenesis.

Entropy production after decoupling of the NLSP and before BBN can
weaken the BBN constraints significantly~\cite{Buchmuller:2006tt}. At
the same time, the gravitino density is diluted which relaxes the
bounds on $T_{\Reheating}$. We show explicitly the effect of entropy
production on the $T_{\Reheating}$ bounds. Here we consider the cases
of late-time entropy production before and after the decoupling of the
NLSP.
Indeed, a relaxation of the $T_{\Reheating}$ bounds can render models
of inflation with $T_{\Reheating} > 10^7~\GeV$ viable in CMSSM
scenarios with gravitino dark matter.
Since also a baryon asymmetry generated in the early Universe is
diluted, the temperature required by thermal leptogenesis increases in
a cosmological scenario with late-time entropy production. Still, we
find that a sufficient amount of entropy production after NLSP
decoupling and before BBN can revive successful thermal leptogenesis.

% _____________________________________________________________________
\section{Thermal gravitino production}
%_____________________________________________________________________

Gravitinos with $\mgr\gtrsim 1~\GeV$ have decoupling temperatures of
$T^{\gravitino}_{\freezeout}\gtrsim 10^{14}~\GeV$, as will be shown
below.
We consider thermal gravitino production in the radiation-dominated
epoch starting at $T_{\Reheating}<T^{\gravitino}_{\freezeout}$
assuming that inflation has diluted away any initial gravitino
population.\footnote{Taking a conservative point of view, we do not
  include gravitino production before the radiation-dominated epoch.
  However, inflaton decays, for example, can lead to a sizable yield
  of non-thermally produced gravitinos depending on the inflation
  model; cf.~\cite{Kawasaki:2006gs,Kawasaki:2006hm} and references
  therein.}
For $T_{\Reheating}<T^{\gravitino}_{\freezeout}$, gravitinos are not
in thermal equilibrium with the post-inflationary plasma. Accordingly,
the evolution of the gravitino number density $n_{\gravitino}$ with
cosmic time $t$ is described by the following Boltzmann
equation~\cite{Pradler:2006qh}
\begin{eqnarray}
&&
    \frac{dn_{\gravitino}}{dt} + 3 H n_{\gravitino} = C_{\gravitino}
\label{Eq:Boltzmann}\\
&&
        C_{\gravitino} 
        =
        \sum_{i=1}^{3} 
        \frac{3\zeta(3)T^6}{16\pi^3\MPl^2} 
        \left(1+\frac{M_i^2}{3\mgr^2}\right)
        c_i\, g_i^2
        \ln\left(\frac{k_i}{g_i}\right)
\label{Eq:CollisionTerm}        
\end{eqnarray}
where $H$ denotes the Hubble parameter. The collision term
$C_{\gravitino}$ involves the gaugino mass parameters $M_i$, the gauge
couplings $g_i$, and the constants $c_i$ and $k_i$ associated with the
gauge groups U(1)$_\Hypercharge$, SU(2)$_\Weak$, and SU(3)$_\Color$ as
given in Table~\ref{Tab:Constants}.
\begin{table}[t]
  \caption{The gauge couplings $g_i$, 
    the gaugino mass parameters $M_i$, and the constants $c_i$, $k_i$, $y_i$, 
    and $\beta_i^{(1)}$ associated with the gauge groups
    U(1)$_\Hypercharge$, SU(2)$_\Weak$, and SU(3)$_\Color$.}
  \label{Tab:Constants}
\begin{center}
\renewcommand{\arraystretch}{1.25}
\begin{tabular*}{3.25in}{@{\extracolsep\fill}cccccccc}
\hline\hline
gauge group         & $i$ & $g_i$ & $M_i$  & $c_i$ &  $k_i$ &  $(y_i/10^{-12})$ & $\beta_i^{(1)}$
\\ \hline
U(1)$_\Hypercharge$ & 1 & $g'$    & $M_1$  & 11    & 1.266  & 0.653 & 11
\\
SU(2)$_\Weak$       & 2 & $g$     & $M_2$  & 27    & 1.312  & 1.604 & 1
\\
SU(3)$_\Color$ & 3 & $g_\mathrm{s}$ & $M_3$ & 72   & 1.271  & 4.276 & -3
\\
\hline\hline
\end{tabular*}
\end{center}
\end{table}
In expression~(\ref{Eq:CollisionTerm}) the temperature $T$ provides
the scale for the evaluation of $M_i$ and $g_i$. The given collision
term is valid for temperatures sufficiently below the gravitino
decoupling temperature, where gravitino disappearance processes can be
neglected. A primordial plasma with the particle content of the
minimal SUSY Standard Model (MSSM) in the high-temperature limit is
used in the derivation of~(\ref{Eq:CollisionTerm}).

The collision term~(\ref{Eq:CollisionTerm}) results from a consistent
gauge-invariant finite-temperature
calculation~\cite{Pradler:2006qh,Pradler:inPreparation} following the
approach used in Ref.~\cite{Bolz:2000fu}. Thus, in contrast to the
previous estimates in~\cite{Ellis:1984eq,Moroi:1993mb}, the expression
for $C_{\gravitino}$ is independent of arbitrary cutoffs. Note that
the field-theoretical methods of~\cite{Braaten:1991dd,Braaten:1989mz}
applied in its derivation require weak couplings, $g_i\ll 1$, and thus
high temperatures $T\gg 10^6~\GeV$.

Assuming conservation of entropy per comoving volume, the Boltzmann
equation~(\ref{Eq:Boltzmann}) can be solved to good approximation
analytically~\cite{Bolz:2000fu,Brandenburg:2004du+X}. At a temperature
$\TL\ll\TR$, the resulting gravitino yield from thermal production
reads
\begin{eqnarray}
        Y_{\gravitino}^{\TP}(\TL)
        &\equiv&
        \frac{n_{\gravitino}^{\TP}(\TL)}{s(\TL)}
        \,\,\approx\,\,
        \frac{C_{\gravitino}(\TR)}{s(\TR)\,H(\TR)}
\nonumber\\
        &=&
        \sum_{i=1}^3
        y_i\, g_i^2(\TR)
        \left(1+\frac{M^2_{i}(\TR)}{3\mgr^2}\right) 
\nonumber\\
        &&\quad\times\ln\left(\frac{k_i}{g_i(\TR)}\right)
        \left(\frac{\TR}{10^{10}\,\GeV} \right)
        \ ,
\label{Eq:YgravitinoTP}
\end{eqnarray}
where the constants $y_i$ are given in Table~\ref{Tab:Constants}. 
These constants are obtained
with the Hubble parameter describing the radiation-dominated epoch, 
$H_{\rad}(T)=\sqrt{g_*(T)\pi^2/90}\, T^2/\MPl$, 
the entropy density
$s(T)=2\pi^2\,g_{*S}(T)\,T^3/45$,
and an effective number of relativistic degrees of freedom of
$g_*(\TR)=g_{*S}(\TR)=228.75$.
We evaluate $g_i(\TR)$ and $M_i(\TR)$ using the one-loop evolution
described by the renormalization group equation in the MSSM:
\begin{eqnarray}
        g_i(T) 
        &=&
        \left(
        g_i^{-2}(m_{\ZB}) - \frac{\beta_i^{(1)}}{8\pi^2}\ln\left[\frac{T}{m_{\ZB}}\right]
        \right)^{-1/2}
        \ ,
\label{Eq:running_coupling}\\
        M_i(T)
        &=&
        \left[\frac{g_i(T)}{g_i(M_{\GUT})}\right]^2 M_i(M_{\GUT})
\label{Eq:running_gaugino_mass}
\end{eqnarray}
with the respective gauge coupling at the Z-boson mass,
$g_i(m_{\ZB})$, and the $\beta_i^{(1)}$ coefficients listed in
Table~\ref{Tab:Constants}.

Without late-time entropy production, the gravitino yield from thermal
production at the present temperature $T_0$ is given by
\begin{equation}
  Y_{\gravitino}^{\TP}(T_0)
  = Y_{\gravitino}^{\TP}(\TL)
  \ .
\label{Eq:YgravitinoTPstandard}
\end{equation}
The resulting density parameter of thermally produced gravitinos is
\begin{equation}
         \Omega_{\gravitino}^{\TP}h^2 
         =  \mgr\,Y_{\gravitino}^{\TP}(T_0)\,s(T_0)\,h^2/\rho_c
\label{Eq:OgravitinoTP}
\end{equation}
with the Hubble constant $h$ in units of
$100~\km\,\Mpc^{-1}\seconds^{-1}$ and
$\rho_c/[s(T_0)h^2]=3.6\times 10^{-9}\,\GeV$.

\begin{figure}[t]
\begin{center}
\includegraphics[width=3.25in]{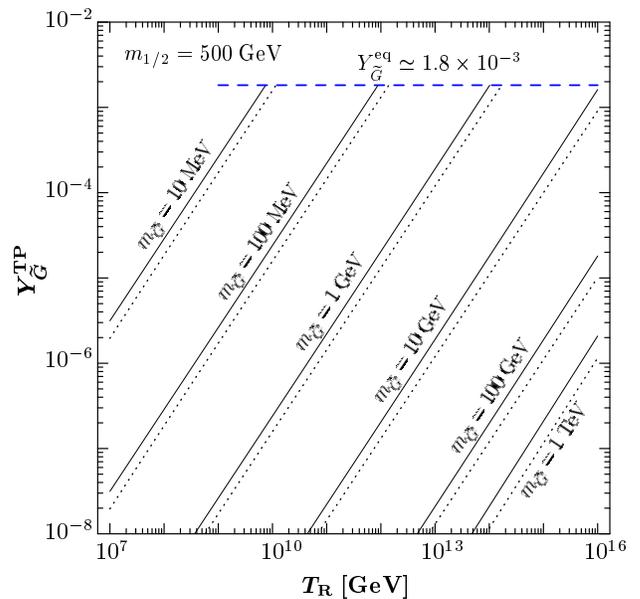} 
\caption{\small The thermally produced gravitino
  yield~(\ref{Eq:YgravitinoTP}) as a function of $T_R$ for
  $\mgr=10~\MeV$, $100~\MeV$, $1~\GeV$, $10~\GeV$, $100~\GeV$, and
  $1~\TeV$ (solid lines from left to right) and
  $M_{1,2,3}(M_{\GUT})=m_{1/2}=500~\GeV$. The dotted lines show the
  corresponding yield obtained with the SU(3)$_\Color$ result for the
  collision term of Ref.~\cite{Bolz:2000fu}. The dashed (blue in the
  web version) horizontal line indicates the equilibrium yield of a
  relativistic spin 1/2 Majorana fermion.}
\label{Fig:YgravitinoTP}
\end{center}
\end{figure}

In Fig.~\ref{Fig:YgravitinoTP} 
our result~(\ref{Eq:YgravitinoTP}) for the thermally produced
gravitino yield $Y_{\gravitino}^{\TP}(\TL)$ is shown as a function of
$T_R$ for various values of $\mgr$ (solid lines).  The curves are
obtained with $m_{1/2}=500~\GeV$ for the case of universal gaugino
masses at $M_{\GUT}$:
$M_{1,2,3}(M_{\GUT})=m_{1/2}$. 
The dotted lines show the corresponding results from the
SU(3)$_\Color$ yield of Ref.~\cite{Bolz:2000fu} for $M_3=m_{1/2}$,
which was used to study $\TR$ constraints on gravitino dark matter
scenarios in
Refs.~\cite{Roszkowski:2004jd,Cerdeno:2005eu,Steffen:2006hw}.  We find
that~(\ref{Eq:YgravitinoTP}) exceeds the yield derived
from~\cite{Bolz:2000fu} by about $50\%$; cf.~\cite{Pradler:2006qh}.
The dashed (blue in the web version) horizontal line indicates the
equilibrium yield
\begin{equation}
  Y_{\gravitino}^{\equil} 
  \equiv \frac{n_{\gravitino}^{\equil}}{s}
  \approx 1.8 \times 10^{-3}
\label{Eq:Y_equil}
\end{equation}
which is given by the equilibrium number density of a relativistic
spin~1/2 Majorana fermion,
$n_{\gravitino}^{\equil} = 3\zeta(3)T^3/(2\pi^2)$.
For $T>T^{\gravitino}_{\freezeout}$,
$g_*(T)=g_{*S}(T)=230.75$
since the spin 1/2 components of the gravitino are in thermal
equilibrium. In the region where the yield~(\ref{Eq:YgravitinoTP})
approaches the equilibrium value~(\ref{Eq:Y_equil}), gravitino
disappearance processes should be taken into account. This would then
lead to a smooth approach of the non-equilibrium yield to the
equilibrium abundance.  Without the backreactions taken into account,
the kink position indicates a lower bound for
$T^{\gravitino}_{\freezeout}$.  Towards smaller $\mgr$,
$T^{\gravitino}_{\freezeout}$ decreases due to the increasing strength
of the gravitino couplings. For example, for $\mgr=1~\GeV$
($10~\MeV$), we find $T^{\gravitino}_{\freezeout}\gtrsim 10^{14}~\GeV$
($10^{10}~\GeV$).

In the analytical expression~(\ref{Eq:YgravitinoTP}) we refer to $\TR$
as the initial temperature of the radiation-dominated epoch. 
So far we have not considered the phase in which the coherent
oscillations of the inflaton field $\phi$ dominate the energy density
of the Universe, where one usually defines $\TR$ in terms of the decay
width~$\Gamma_{\phi}$ of the inflaton field~$\phi$.
To account for the reheating phase, we numerically
integrate~(\ref{Eq:Boltzmann}) together with the Boltzmann equations
for the energy densities of radiation and the inflaton field,
\begin{eqnarray}
  \frac{d\rho_{\rad}}{dt} + 4 H \rho_{\rad} 
  &=&
  \Gamma_{\phi} \rho_{\phi} \ ,
\label{Eq:BEqPhi}
  \\
  \frac{d\rho_{\phi}}{dt} + 3 H \rho_{\phi}
  &=&
  -\Gamma_{\phi} \rho_{\phi} \ ,
\label{Eq:BEqRad}
\end{eqnarray}
respectively; for details see Appendix~F of Ref.~\cite{Kawasaki:2004qu}.

With our result for the collision term~(\ref{Eq:CollisionTerm}), we
find that the gravitino yield obtained numerically is in good
agreement with the analytical expression~(\ref{Eq:YgravitinoTP}) for
\begin{equation}
  \TR 
  \simeq 
  \left[\frac{90}{g_*(\TR)\pi^2}\right]^{1/4}
  \sqrt{\frac{\Gamma_{\phi}\,\MPl}{1.8}}
\label{Eq:TR_definition}
\end{equation}
which satisfies $\Gamma_{\phi}\simeq 1.8\,H_{\rad}(\TR)$. For an
alternative $\TR$ definition
given by
$\Gamma_{\phi}=\xi\,H_{\rad}(\TR)$, 
\begin{equation}
  \TR^{\xi} 
  \equiv
  \left[\frac{90}{g_*(\TR)\pi^2}\right]^{1/4}
  \sqrt{\frac{\Gamma_{\phi}\,\MPl}{\xi}}
  \ ,
\label{Eq:TRxi_definition}
\end{equation}
the associated numerically obtained gravitino yield is described by
the analytical expression obtained after substituting $\TR$ with
$\sqrt{\xi/1.8}\,\TR^{\xi}$ in~(\ref{Eq:YgravitinoTP}).

While we focus on scenarios in which the gravitino is stable, the
yield~(\ref{Eq:YgravitinoTP}) is also crucial to extract cosmological
constraints in scenarios with unstable gravitinos. Based on the result
of~\cite{Bolz:2000fu} and taking into account thermal gravitino
production during reheating, the following fitting
formula was used to study constraints from decaying gravitinos in
Refs.~\cite{Kawasaki:2004yh,Kawasaki:2004qu,Kohri:2005wn}:
\begin{eqnarray}
    Y_{\gravitino}^{\mathrm{KKM}}(\TL)
    &\simeq&
    1.9\times 10^{-12} \left(\frac{\TR}{10^{10}~\GeV}\right) 
\nonumber\\
    && \!\!\times\left[1+0.045\,\ln\left(\frac{\TR}{10^{10}~\GeV}\right)\right]            
\nonumber\\
    && \!\!\times\left[1-0.028\,\ln\left(\frac{\TR}{10^{10}~\GeV}\right)\right]\!,        
\label{Eq:Y_KKM}
\end{eqnarray}
where $\TR$ was defined via
$\Gamma_{\phi}=3\, H_{\rad}(\TR)$. 
Comparing~(\ref{Eq:Y_KKM}) with our result after the matching of the
$\TR$ definitions, we find that our result exceeds the
$\mgr$-independent yield~(\ref{Eq:Y_KKM}) by about 30\% for $\mgr \gg
M_i(\TR)$.  While the $\mgr$ dependence of~$Y_{\gravitino}^{\TP}$
becomes negligible for decreasing $M_i(\TR)/\mgr$, the
yield~(\ref{Eq:Y_KKM}) is used for $\mgr$ as small as $100~\GeV$ in
Refs.~\cite{Kawasaki:2004yh,Kawasaki:2004qu,Kohri:2005wn}.  As can be
seen in Fig.~\ref{Fig:YgravitinoTP}, the actual yield for
$\mgr=100~\GeV$ is thereby underestimated by about an order of
magnitude.  Accordingly, the $\TR$ bounds given
in~\cite{Kawasaki:2004yh,Kawasaki:2004qu,Kohri:2005wn} are
underestimated in the region $\mgr<1~\TeV$.

%_____________________________________________________________________
\section{Constraints on \boldmath$T_{\Reheating}$}
%_____________________________________________________________________

The reheating temperature $T_{\Reheating}$ is limited from above in
the case of a stable gravitino LSP since $\Omega_{\gravitino}^{\TP}$
cannot exceed the dark matter density
$\Omega_{\CDM}$~\cite{Moroi:1993mb,Asaka:2000zh,Roszkowski:2004jd,Cerdeno:2005eu,Steffen:2006hw}.
In this paper, we use~\cite{Spergel:2006hy,PDB2006}
\begin{equation}
        \Omega_{\CDM}^{3\sigma}h^2=0.105^{+0.021}_{-0.030} 
\label{Eq:OmegaDM}
\end{equation} 
as obtained from the measurements of the cosmic microwave background
(CMB) anisotropies by the Wilkinson Microwave Anisotropy Probe (WMAP)
satellite.\footnote{This nominal $3\sigma$ range is derived assuming a
  restrictive six-parameter ``vanilla'' model. A larger range is
  possible---even with additional data from other cosmological
  probes---if the fit is performed in the context of a more general
  model~\cite{Hamann:2006pf}.}

\begin{figure}[t]
\begin{center}
\includegraphics[width=3.25in]{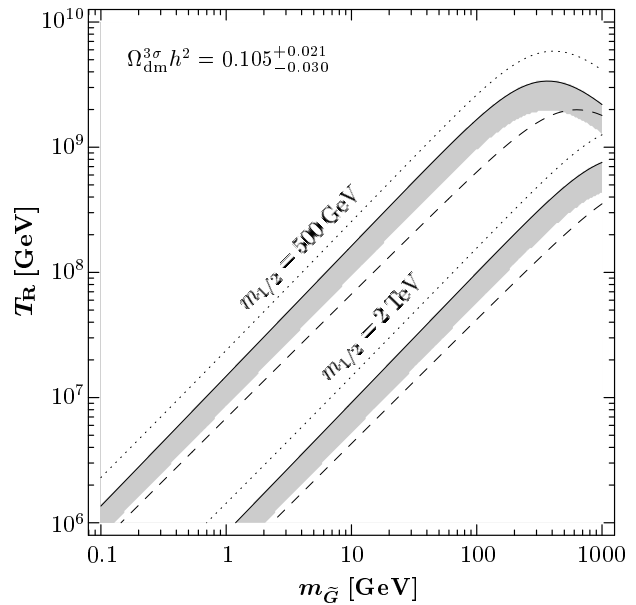} 
\caption{\small Upper limits on the reheating temperature
  $T_{\Reheating}$. On the upper (lower) gray band,
  $\Omega_{\widetilde{G}}^{\TP}$ for $M_{1,2,3}=m_{1/2}=500~\GeV$
  ($2~\TeV$) at $M_{\GUT}$ agrees with $\Omega_{\CDM}^{3\sigma}$. The
  corresponding $\TR$ limits from the requirement
  $\Omega_{\widetilde{G}}^{\TP}h^2\leq 0.126$ shown by the dashed and
  dotted lines are obtained respectively with~(\ref{Eq:YgravitinoTP})
  for $M_1/10=M_2/2=M_3=m_{1/2}$ at $M_{\GUT}$ and with the result of
  Ref.~\cite{Bolz:2000fu} for $M_3=m_{1/2}$ at $M_{\GUT}$.}
\label{Fig:UpperLimitTR}
\end{center}
\end{figure}

In Fig.~\ref{Fig:UpperLimitTR}
we show the resulting upper limits on $T_{\Reheating}$ as a function
of $\mgr$. On the gray band, the thermally produced gravitino
density~(\ref{Eq:OgravitinoTP}) is within the nominal $3\sigma$
range~(\ref{Eq:OmegaDM}). The upper (lower) gray band is obtained for
$M_{1,2,3}=m_{1/2}$ at $M_{\GUT}$ with $m_{1/2}=500~\GeV$ ($2~\TeV$).
The dashed lines show the corresponding constraints for the exemplary
non-universal scenario~\cite{Anderson:1996bg}
$M_1/10=M_2/2=M_3=m_{1/2}$ at $M_{\GUT}$ and from the requirement
$\Omega_{\widetilde{G}}^{\TP}h^2\leq 0.126$.  Using the same
requirement and the result of~\cite{Bolz:2000fu} for $M_3=m_{1/2}$, we
obtain the dotted lines.
The electroweak contributions are particularly important for the
considered case of non-universal gaugino masses at $M_{\GUT}$. For
universal gaugino masses at $M_{\GUT}$, the $\TR$ limits derived with
the result of~\cite{Bolz:2000fu} provide already reasonable
estimates.\footnote{Note that the dotted curve shown for
  $m_{1/2}=500~\GeV$ in Fig.~\ref{Fig:UpperLimitTR} is by about a
  factor of~4 more severe than the $\TR$ limits shown in Figs.~5 and~6
  of Ref.~\cite{Cerdeno:2005eu} in the region $\mgr\lesssim 10~\GeV$ in
  which $\Omega_{\widetilde{G}}^{\TP}$ governs the limits.
  It seems to us that the gluino mass in (1.2) of
  Ref.~\cite{Cerdeno:2005eu} was accidentally evaluated at the scale
  $\mu=\TR$ rather than at the scale $\mu\simeq 100~\GeV$; see Sec.~5
  of~\cite{Bolz:2000fu}.}

The $\TR$ limits shown in Fig.~\ref{Fig:UpperLimitTR} are conservative
bounds that do only depend on $\mgr$ and the $M_i$ values at
$M_{\GUT}$. Once details of the SUSY model realized in Nature are
known, one will be able to refine the limits by including
contributions to $\Omega_{\widetilde{G}}$ from NLSP decays.  In the
next section, we will account for this non-thermal gravitino
production in a systematic way within the framework of the CMSSM.

%_____________________________________________________________________
\section{Constraints on \boldmath$T_{\Reheating}$ in the CMSSM}
%_____________________________________________________________________

In the CMSSM, one assumes universal soft SUSY breaking parameters at
$M_{\GUT}$. The CMSSM yields phenomenologically acceptable spectra
with only four parameters and a sign: the gaugino mass parameter
$m_{1/2}$, the scalar mass parameter $m_0$, the trilinear coupling
$A_0$, the mixing angle $\tan\beta$ in the Higgs sector, and the sign
of the higgsino mass parameter $\mu$.

Assuming $A_0=0$ for simplicity, the lightest Standard Model
superpartner is either the lightest neutralino $\neutralino$ or the
lighter stau $\stau$. Indeed, most CMSSM investigations assume that
$\neutralino$ is the LSP that provides dark matter;
cf.~\cite{Djouadi:2006be} and references therein. The parameter region
in which $m_{\stau}<m_{\neutralino}$ is usually not considered because
of the severe upper limits on the abundance of stable charged
particles~\cite{PDB2006}. In the gravitino LSP case,
$m_{\stau}<m_{\neutralino}$ can be viable since the lightest Standard
Model superpartner is
unstable~\cite{Ellis:2003dn,Roszkowski:2004jd,Cerdeno:2005eu,Cyburt:2006uv}.

With the gravitino LSP, the lightest Standard Model superpartner is
the NLSP that decays into Standard Model particles and one gravitino
LSP. For $\mgr\gtrsim 1~\GeV$, the NLSP decays after its decoupling
from the thermal plasma.\footnote{The NLSP freezeout temperature can
  be estimated from its mass: $T_{\freezeout}^{\NLSP}\lesssim
  m_{\NLSP}/20$~\cite{Asaka:2000zh}.  Thus, $\TR\gg
  T_{\freezeout}^{\NLSP}$ for $\TR> 10^6~\GeV$ which is considered in
  this Letter.} Thus, the relic density of the associated
non-thermally produced gravitinos reads~\cite{Borgani:1996ag}
\begin{eqnarray}
        \Omega_{\widetilde{G}}^{\NTP}h^2
        &=& \mgr\,Y_{\NLSP}(T_0)\,s(T_0)\,h^2/\rho_c
\label{Eq:OgravitinoNTP}
\\
        &=& \frac{\mgr}{m_{\NLSP}}\,\Omega_{\NLSP}h^2
        \ ,
\label{Eq:GravitinoDensityNTP}
\end{eqnarray}
where $m_{\NLSP}$ is the mass of the NLSP and $Y_{\NLSP}(T_0)$ and
$\Omega_{\NLSP}h^2$ are respectively the yield and the relic density
that the NLSP would have today, if it had not decayed. 

\begin{figure*}[t]
\begin{center}
\includegraphics[width=3.25in]{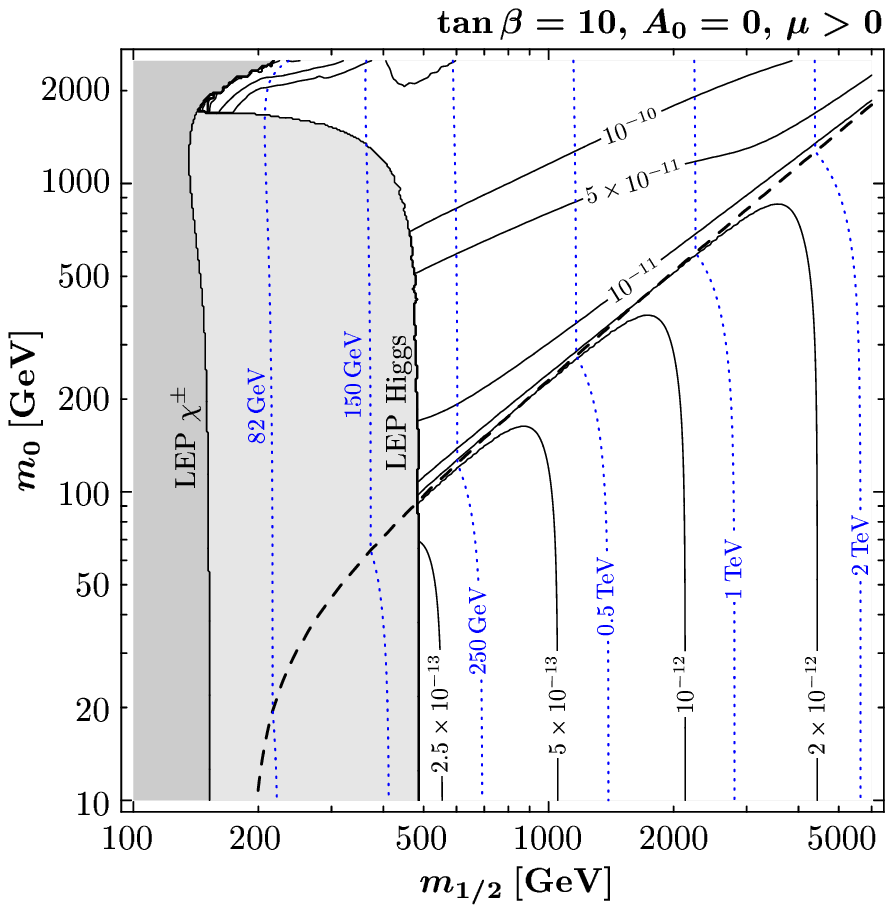} 
\hskip 0.75cm
\includegraphics[width=3.25in]{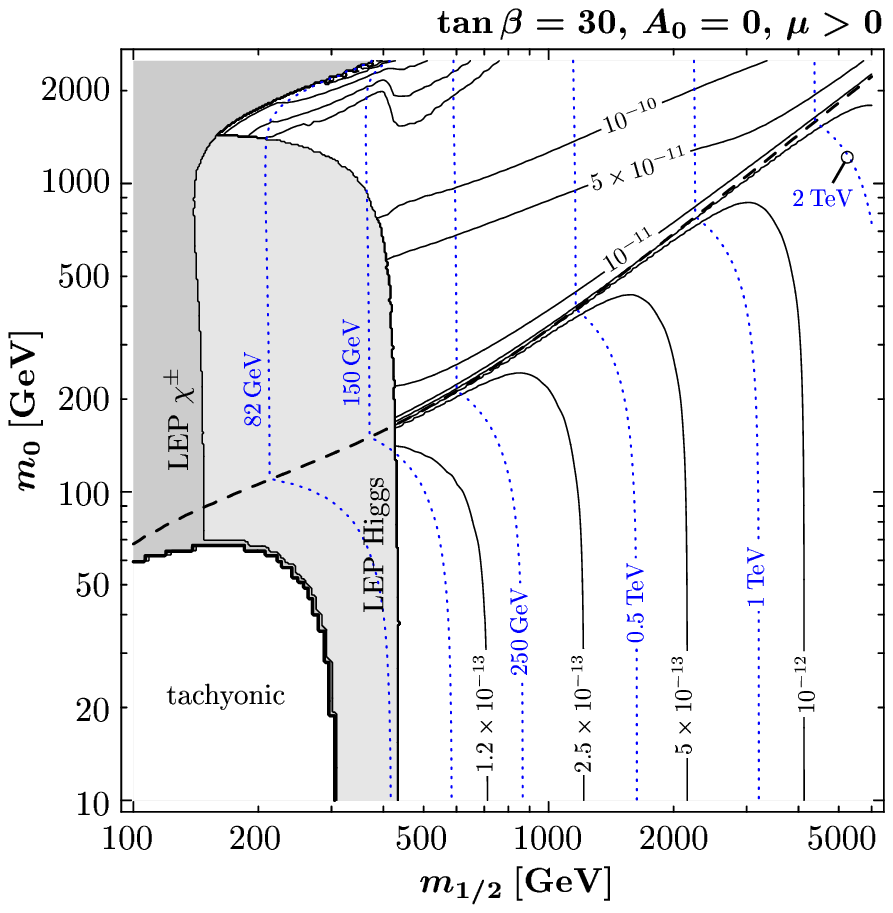} 
\caption{\small Contours of $Y_{\NLSP}(T_0)$ (solid black lines) and
  $m_{\NLSP}$ (dotted blue lines) in the $(m_{1/2},m_0)$ plane for
  $A_0=0$, $\mu>0$, $\tan\beta=10$ (left panel) and $\tan\beta=30$
  (right panel). Above (below) the dashed line,
  $m_{\neutralino}<m_{\stau}$ ($m_{\stau}<m_{\neutralino}$). The
  medium gray and the light gray regions at small $m_{1/2}$ show the
  mass bounds $m_{\chargino}>94~\GeV$ and $m_{\Higgs}>114.4~\GeV$ from
  chargino and Higgs searches at LEP~\cite{PDB2006}.}
\label{Fig:YNLSP}
\end{center}
\end{figure*}

In Fig.~\ref{Fig:YNLSP}
the solid (black) and dotted (blue in the web version) lines show
respectively contours of $Y_{\NLSP}(T_0)$ and $m_{\NLSP}$ in the
$(m_{1/2},m_0)$ plane for $A_0=0$, $\mu>0$, $\tan\beta=10$ (left
panel) and $\tan\beta=30$ (right panel). Above (below) the dashed
line, $m_{\neutralino}<m_{\stau}$ ($m_{\stau}<m_{\neutralino}$). The
medium gray and the light gray regions at small $m_{1/2}$ are excluded
respectively by the mass bounds $m_{\chargino}>94~\GeV$ and
$m_{\Higgs}>114.4~\GeV$ from chargino and Higgs searches at
LEP~\cite{PDB2006}. The leftmost dotted (blue in the web version) line
indicates the LEP bound $m_{\stau}>81.9~\GeV$~\cite{PDB2006}.
For $\tan\beta=30$, tachyonic sfermions occur in the low-energy
spectrum at points in the white corner labeled as ``tachyonic.''
We employ the \texttt{FORTRAN} program
\texttt{SuSpect}~\cite{Djouadi:2002ze} to calculate the low-energy
spectrum of the superparticles and the Higgs bosons, where we use
$m_{\mathrm{t}}=172.5~\GeV$ for the top quark mass. Assuming standard
cosmology, the yield $Y_{\NLSP}(T_0)$ is obtained from the
$\Omega_{\NLSP}h^2$ values provided by the computer program
\texttt{micrOMEGAs}~\cite{Belanger:2001fz+X}.

The contours shown in Fig.~\ref{Fig:YNLSP} are independent of $\mgr$
and $\TR$. Therefore, they can be used to interpret the results shown
in the figures below. Note the sensitivity of both $Y_{\stau}(T_0)$
and $m_{\stau}$ on $\tan\beta$.  By going from $\tan\beta=10$ to
$\tan\beta=30$, $Y_{\stau}(T_0)$ decreases by about a factor of two at
points that are not in the vicinity of the dashed line, i.e., that are
outside of the $\stau$--$\neutralino$ coannihilation region. While
$m_{\stau}$ becomes smaller by increasing $\tan\beta$ to 30, the
$\tan\beta$ dependence of $m_{\neutralino}$ is negligible.

Let us now explore the parameter space in which
\begin{equation}
        0.075
        \leq 
        \Omega_{\gravitino}^{\TP}h^2+\Omega_{\gravitino}^{\NTP}h^2 
        \leq 
        0.126
        \ .
\label{Eq:OgravitinoConstraint}
\end{equation} 
Now, $\TR$ and $\mgr$ appear in addition to the traditional CMSSM
parameters. We focus on $\mgr\gtrsim 1~\GeV$ since the soft SUSY
breaking parameters of the CMSSM are usually assumed to result from
gravity-mediated SUSY breaking. However, we do not restrict our study
to fixed relations between $\mgr$ and the soft SUSY breaking
parameters such as the ones suggested, for example, by the Polonyi
model.

\begin{figure*}[t]
\begin{center}
\includegraphics[width=3.25in]{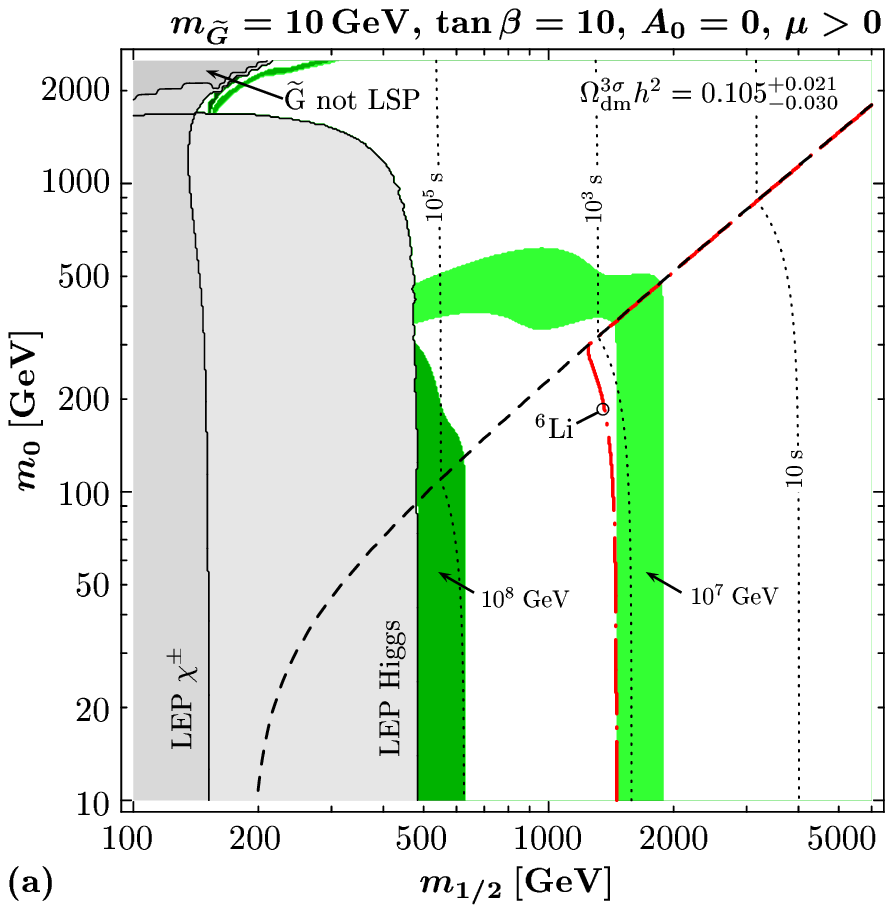} 
\hskip 0.75cm
\includegraphics[width=3.25in]{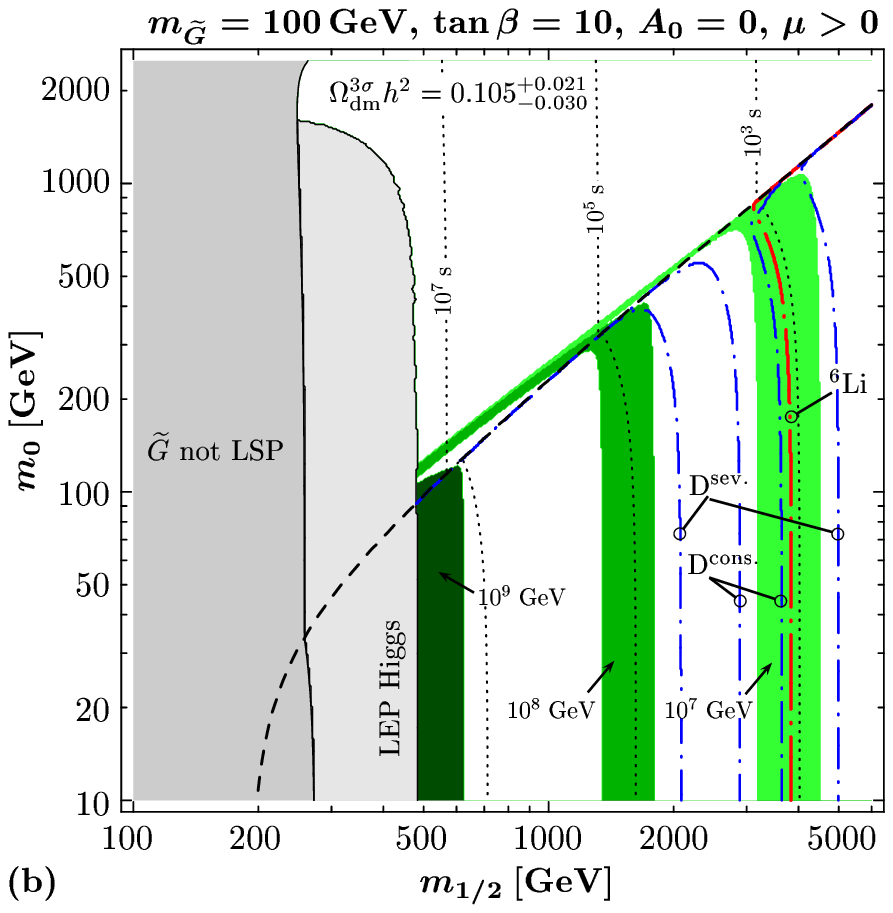} 
\vskip 0.75cm
\includegraphics[width=3.25in]{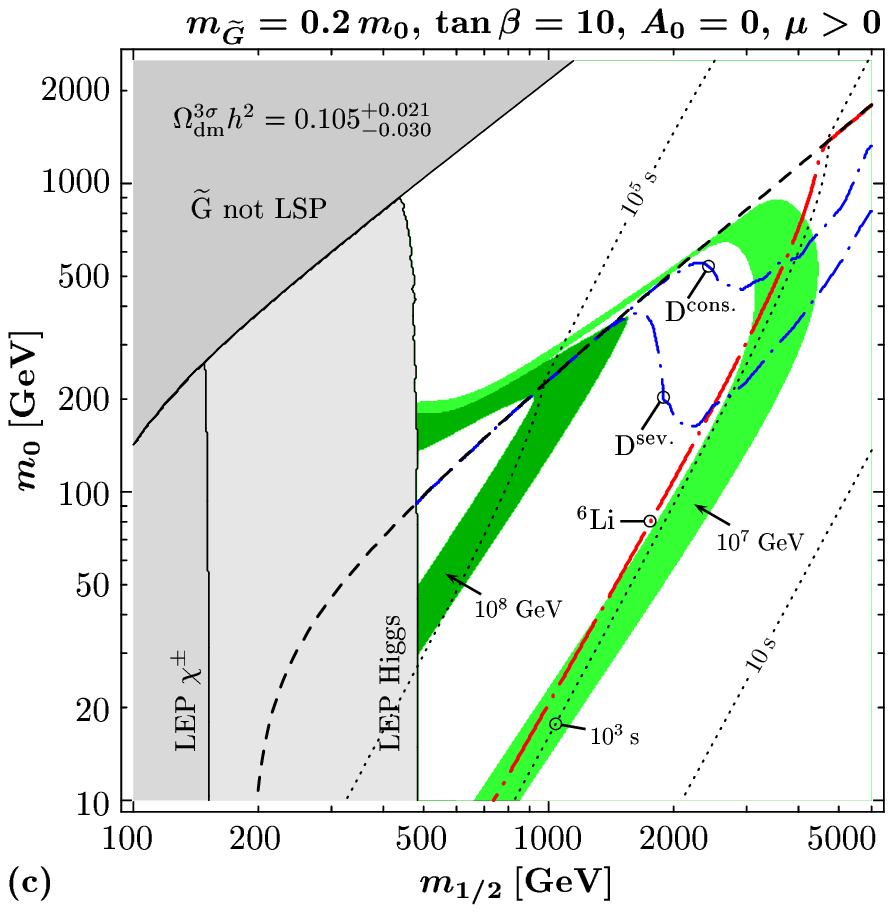} 
\hskip 0.75cm
\includegraphics[width=3.25in]{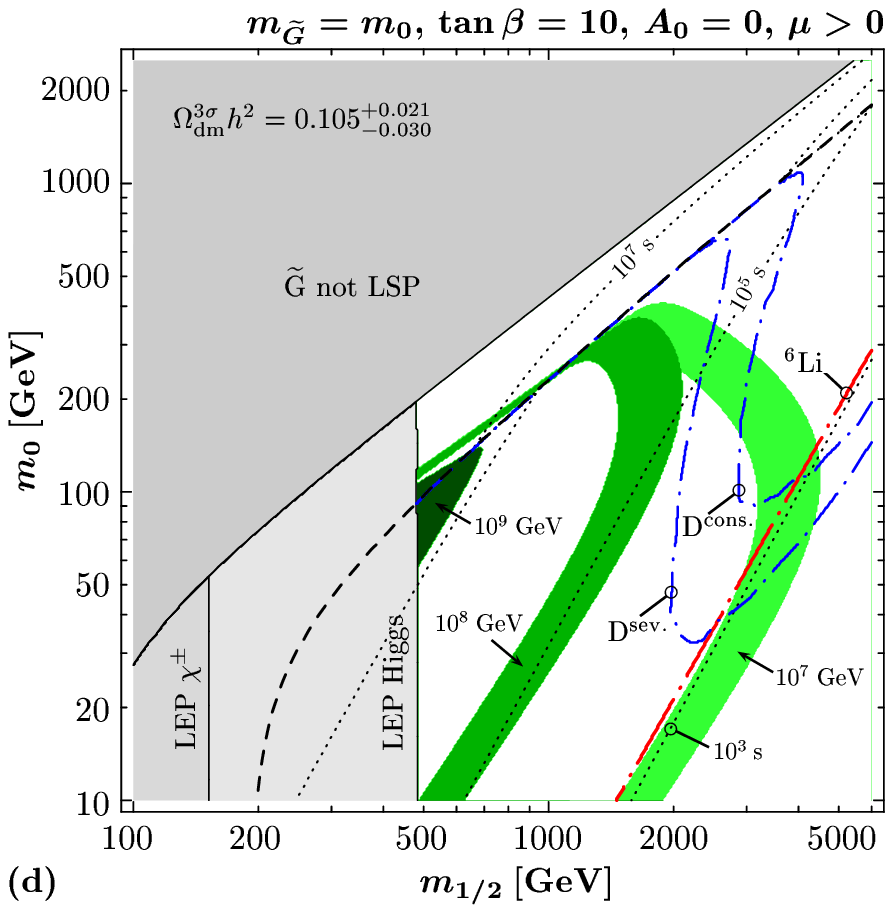} 
\caption{\small The $(m_{1/2},m_0)$ planes for $\tan\beta=10$,
  $A_0=0$, $\mu>0$, and the choices (a)~$\mgr=10~\GeV$,
  (b)~$\mgr=100~\GeV$, (c)~$\mgr=0.2\,m_0$, and (d)~$\mgr=m_0$.  In
  each panel, the light, medium, and dark shaded (green in the web version)
  bands indicate the regions in which
  $0.075\leq\Omega_{\gravitino}h^2\leq 0.126$ for $\TR=10^7$, $10^8$, and
  $10^9~\GeV$, respectively. 
  The medium gray and the light gray regions at small $m_{1/2}$ are
  excluded respectively by chargino and Higgs searches at LEP.  In the
  dark gray region, the gravitino is not the LSP. The dotted lines
  show contours of the NLSP lifetime. Below the dashed line,
  $m_{\stau}<m_{\neutralino}$. With the $\stau$ NLSP, the region to
  the left of the long-dash-dotted (red in the web version) line is
  cosmologically disfavored by bound-state effects on the primordial $^6$Li
  abundance~\cite{Pospelov:2006sc}. 
The effects of late hadronic energy injection on the primordial D abundance~\cite{Steffen:2006hw} 
disfavor the $\stau$ NLSP region between the
short-dash-dotted (blue in the web version) lines in panel~(b) and the one
above the corresponding lines in panels~(c) and~(d).
The $\neutralino$ NLSP region above the dashed line, in which
$m_{\neutralino}<m_{\stau}$, is cosmologically disfavored by the
effects of late electromagnetic/hadronic energy injection on the
abundances of the light primordial
elements~\cite{Ellis:2003dn,Feng:2004mt,Roszkowski:2004jd,Cerdeno:2005eu,Cyburt:2006uv}.}
\label{Fig:CMSSMtB10}
\end{center}
\end{figure*}

In Fig.~\ref{Fig:CMSSMtB10}
the light, medium, and dark shaded (green in the web version) bands
show the $(m_{1/2},m_0)$ regions that satisfy the
constraint~(\ref{Eq:OgravitinoConstraint}) for $\TR=10^7$, $10^8$, and
$10^9~\GeV$, respectively, where $\tan\beta=10$, $A_0=0$, $\mu>0$. The
four panels are obtained for the choices (a)~$\mgr=10~\GeV$,
(b)~$\mgr=100~\GeV$, (c)~$\mgr=0.2\,m_0$, and (d)~$\mgr=m_0$. In the
dark-gray region, the gravitino is not the LSP. The regions excluded
by the chargino and Higgs mass bounds and the line indicating
$m_{\neutralino}=m_{\stau}$ are identical to the ones shown in the
left panel of Fig.~\ref{Fig:YNLSP}.  The dotted lines show contours of
the NLSP lifetime. For the $\stau$ NLSP,
\begin{equation}
        \tau_{\st} 
        \simeq \Gamma^{-1}(\stau\to\gravitino\tau)
        = \frac{48 \pi \mgr^2 \MPl^2}{m_{\st}^5} 
        \left(1-\frac{\mgr^2}{m_{\st}^2}\right)^{-4}
\label{Eq:StauLifetime}
\end{equation}
as obtained in the limit $m_{\tau}\to 0$. 
For the $\neutralino$ NLSP, we calculate $\tau_{\neutralino}$ from the
expressions given in Sec.~IIC of Ref.~\cite{Feng:2004mt}.

The $\tau_{\NLSP}$ contours in Fig.~\ref{Fig:CMSSMtB10} illustrate
that the NLSP decays during/after BBN. Successful BBN predictions
therefore imply cosmological constraints on $\mgr$, $m_{\NLSP}$, and
$Y_{\NLSP}$~\cite{Feng:2004mt,Roszkowski:2004jd,Cerdeno:2005eu,Steffen:2006hw}.
Indeed, it has been found that the considered $\neutralino$ NLSP
region is completely disfavored for $\mgr\gtrsim 1~\GeV$ by
constraints from late electromagnetic and hadronic energy
injection~\cite{Feng:2004mt,Roszkowski:2004jd,Cerdeno:2005eu,Cyburt:2006uv}.
In the $\stau$ NLSP region, the constraints from electromagnetic and
hadronic energy release are important but far less severe than in the
$\neutralino$ NLSP case. Thus, much of the $\stau$ NLSP region was
believed to be cosmologically
allowed~\cite{Feng:2004mt,Roszkowski:2004jd,Cerdeno:2005eu,Steffen:2006hw}.

Recently, this picture has changed. It has been found that bound-state
formation of long-lived negatively charged $\stau$'s with primordial
nuclei can catalyze the production of $^6$Li
significantly~\cite{Pospelov:2006sc,Cyburt:2006uv}. Indeed, in most of
the $\stau$ NLSP parameter space, the associated bounds are much more
severe than the ones from late energy injection.
Only for $\tau_{\st}\lesssim 10^3~\seconds$ and $\mgr\gtrsim 40~\GeV$,
the constraints from hadronic energy release can become more severe
than the ones from catalyzed $^6$Li
production~\cite{Cyburt:2006uv,Steffen:2006wx}.
We thus consider both the constraint from catalyzed $^6$Li production
derived in~\cite{Pospelov:2006sc} and the one from late hadronic
energy injection derived in~\cite{Steffen:2006hw}.\footnote{For
  details on the other BBN bounds and the additional CMB bounds, we
  refer the reader to the detailed investigations presented in
  Refs.~\cite{Ellis:2003dn,Feng:2004mt,Roszkowski:2004jd,Cerdeno:2005eu,Lamon:2005jc}.}

For the constraint from bound-state effects on $^6$Li production, we
adopt the bounds given in Fig.~4 of Ref.~\cite{Pospelov:2006sc} as
$\tau_{\st}$-dependent upper limits on the yield of the negatively
charged staus, $Y_{\NLSP}/2$. These bounds are obtained assuming a
limiting primordial abundance of~\cite{Cyburt:2002uv}
\begin{equation}
(^6\mathrm{Li/H})_{\mathrm{p}}\lesssim 2\times 10^{-11}
\ . 
\label{Eq:LiAbundance}
\end{equation}
The resulting constraint disfavors the $\stau$ NLSP region to the left
of the long-dash-dotted (red in the web version) line shown in
Fig.~\ref{Fig:CMSSMtB10}.

For the constraint from late hadronic energy injection, we use the
upper limits on $Y_{\NLSP}$ that are given in Fig.~11 of
Ref.~\cite{Steffen:2006hw}.
These limits are derived from a computation of the 4-body decay of the
stau NLSP into the gravitino, the tau, and a quark-antiquark
pair.\footnote{The 3-body estimate of the hadronic energy release
  given in Ref.~\cite{Feng:2004mt} leads to overly restrictive limits,
  as shown in Ref.~\cite{Steffen:2006hw}.}
They are based on the severe and conservative upper bounds on the
released hadronic energy (95\% CL) obtained in~\cite{Kawasaki:2004qu}
for observed values of the primordial D abundance of
\begin{eqnarray}
(n_{\mathrm{D}}/n_{\mathrm{H}})_{\mathrm{mean}} 
&=& (2.78^{+0.44}_{-0.38})\times 10^{-5} 
\quad \mathrm{(severe)},
\nonumber\\
(n_{\mathrm{D}}/n_{\mathrm{H}})_{\mathrm{high}} 
&=& (3.98^{+0.59}_{-0.67})\times 10^{-5}
\quad \mathrm{(conservative)}. 
\nonumber
\end{eqnarray}
In Fig.~\ref{Fig:CMSSMtB10} the associated constraints are shown by
the short-dash-dotted (blue in the web version) lines. The D
constraint disfavors the region between the corresponding lines in
panel~(b) and the region above the corresponding lines in panels~(c)
and~(d).  In panel~(a) the D constraint does not
appear.\footnote{Additional constraints on hadronic energy release are
  imposed by the primordial abundances of $^4$He, $^3$He/D, $^7$Li,
  and
  $^6$Li/$^7$Li~\cite{Sigl:1995kk,Jedamzik:1999di,Jedamzik:2004er,Kawasaki:2004qu,Jedamzik:2006xz,Cyburt:2006uv}.
  However, in the region allowed by the $^6$Li constraint from
  bound-state effects, i.e., $\tau_{\st}\lesssim 10^3~\seconds$, the
  considered D constraint on hadronic energy release is the dominant
  one as can be seen in Figs.~38--41 of Ref.~\cite{Kawasaki:2004qu}
  and Figs.~6--8 of Ref.~\cite{Jedamzik:2006xz}.}

Remarkably, one finds in each panel of Fig.~\ref{Fig:CMSSMtB10} that
the highest $\TR$ value allowed by the considered BBN constraints is
about $10^7~\GeV$. The bands obtained for $\TR\gtrsim 10^8~\GeV$ are
located completely within the region disfavored by the $^6$Li bound.
In previous gravitino dark matter studies within the CMSSM
that did not take into account bound-state effects on the primordial
$^6$Li abundance,
much higher temperatures of up to about $10^9~\GeV$ were believed to
be
allowed~\cite{Roszkowski:2004jd,Cerdeno:2005eu,Pradler:2006qh}.\footnote{Note
  that our bands for $\TR=10^9~\GeV$ differ from the ones shown in
  Refs.~\cite{Roszkowski:2004jd,Cerdeno:2005eu}; see footnote~4.}

\begin{figure}[t]
\begin{center}
\includegraphics[width=3.25in]{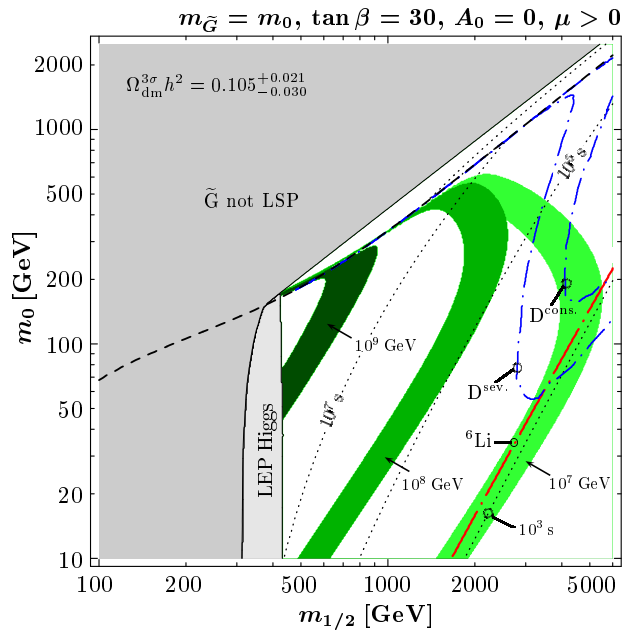} 
\caption{\small As in Fig.~\ref{Fig:CMSSMtB10}~(d), for
  $\tan\beta=30$, $A_0=0$, $\mu>0$, and $\mgr=m_0$.}
\label{Fig:CMSSMtB30}
\end{center}
\end{figure}

The constraint $\TR\lesssim 10^7~\GeV$ remains if we consider larger
values of $\tan\beta$. This is demonstrated in
Fig.~\ref{Fig:CMSSMtB30} for $\tan\beta=30$, $A_0=0$, $\mu>0$, and
$\mgr=m_0$.
The shadings (colors in the web version) and line styles are identical
to the ones in Fig.~\ref{Fig:CMSSMtB10}.

Let us comment on the dependence of the considered BBN constraints on
the assumed primordial abundances of D and $^6$Li. As can be seen in
Figs.~\ref{Fig:CMSSMtB10} and~\ref{Fig:CMSSMtB30}, the constraint from
late hadronic energy release is quite sensitive on the assumed
primordial D abundance. In contrast, even if we relax the restrictive
$^6$Li bound on $Y_{\NLSP}/2$ by two orders of magnitude, we still
find $\TR\lesssim 10^7~\GeV$. For example, the $^6$Li constraint
relaxed in this way would appear in Fig.~\ref{Fig:CMSSMtB10}~(b) as an
almost vertical line slightly above $m_{1/2}=3~\TeV$.

While the constraint $\TR\lesssim 10^7~\GeV$ is found for each of the
considered $\mgr$ relations, one cannot use the $^6$Li bound to set
bounds on $m_{\stau}$ without insights into $\mgr$. The $^6$Li bound
disappears for
$\tau_{\stau}\lesssim 10^3~\seconds$~\cite{Pospelov:2006sc}
which is possible even for $m_{\stau}=\Order(100~\GeV)$ provided
$\mgr$ is sufficiently small; see~(\ref{Eq:StauLifetime}). However,
the constraints on $\TR$ become more severe towards small $\mgr$ as is
shown in Fig.~\ref{Fig:UpperLimitTR}. Thus, the constraint $\TR\lesssim
10^7~\GeV$ cannot be evaded by lowering $\mgr$ provided
$\TR<T_{\freezeout}^{\gravitino}$. 

An upper limit on $\TR$ of $10^7~\GeV$ can be problematic for
inflation models and baryogenesis scenarios. This finding can thus be
important for our understanding of the thermal history of the
Universe.

%_____________________________________________________________________
\section{Constraints on \boldmath$\TR$ with late-time entropy
  production}
%_____________________________________________________________________

The constraints shown above are applicable for a standard thermal
history during the radiation-dominated epoch.  However, it is possible
that a substantial amount of entropy is released, for example, in
out-of-equilibrium decays of a long-lived massive particle
species~X~\cite{Scherrer:1984fd,Kolb:vq}.\footnote{Gravitino dark
  matter scenarios with late-time entropy production have been
  considered previously for gauge-mediated SUSY breaking
  where~$\TR>T^{\gravitino}_{\freezeout}$~\cite{Baltz:2001rq,Fujii:2002fv+X,Jedamzik:2005ir+X}.}

If X lives sufficiently long, it might decay while its rest mass
dominates the energy density of the Universe.  The associated
evolution of the entropy per comoving volume, $S\equiv s\,a^3$, is
described by~\cite{Scherrer:1984fd,Kolb:vq}
\begin{equation}
        \frac{dS}{dt}
        =\frac{\Gamma_{\mathrm{X}}\rho_{\mathrm{X}} a^3}{T}
%        S^{1/3}\frac{dS}{dt}
        =\left(\frac{2\pi^2}{45}\,g_{*}\right)^{1/3} \Gamma_{\mathrm{X}}\rho_{\mathrm{X}} a^4 S^{-1/3}
\label{Eq:EntropyEvolution}
\end{equation}
together with the Boltzmann equation~(\ref{Eq:BEqRad}) for
$\phi=\mathrm{X}$ and the Friedmann equation governing the evolution
of the scale factor of the Universe $a$. Here $\Gamma_{\mathrm{X}}$
and $\rho_{\mathrm{X}}$ denote respectively the decay width and the
energy density of X.
Thus, the temperature after the decay can be expressed in terms of
$\Gamma_{\mathrm{X}}$,
\begin{equation}
  T_{\mathrm{after}} 
  \equiv
  \left[\frac{10}{g_*(T_{\mathrm{after}})\pi^2}\right]^{1/4}
  \sqrt{\Gamma_{\mathrm{X}}\,\MPl}
  \ ,
\label{Eq:Tafter_definition}
\end{equation}
which satisfies $\Gamma_{\mathrm{X}}=3 H_{\rad}(T_{\mathrm{after}})$.
Indeed, primordial nucleosynthesis imposes a lower limit on this
temperature~\cite{Kawasaki:1999na,Kawasaki:2000en,Hannestad:2004px,Ichikawa:2005vw}:
\begin{equation}
        T_{\mathrm{after}}\gtrsim 0.7\!-\!4~\MeV
        \ .
\label{Eq:Tafter_limit}
\end{equation}

\begin{figure}[t]
\begin{center}
\includegraphics[width=3.25in]{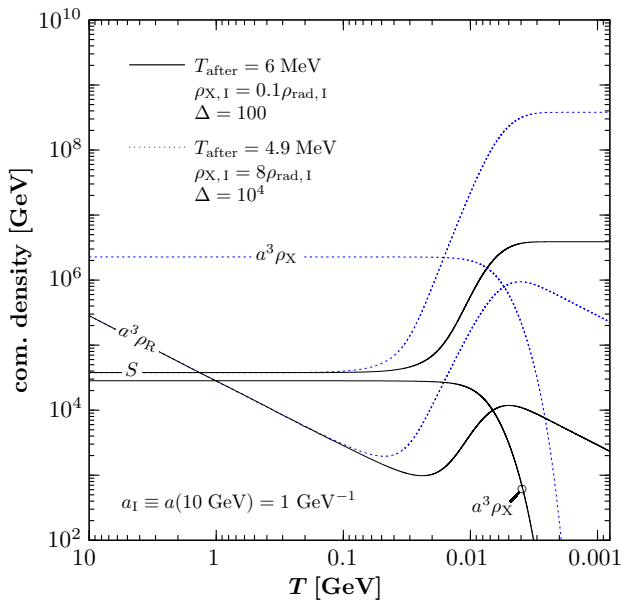} 
\caption{\small Evolution of $S$, $a^3\rho_{\mathrm{X}}$, and
  $a^3\rho_{\rad}$ as a function of $T$ for the normalization
  $a_{\mathrm{I}}\equiv a(10~\GeV)=1~\GeV^{-1}$.  The solid lines are
  obtained for $\rho_{\mathrm{X}}(10~\GeV)=0.1\,\rho_{\rad}(10~\GeV)$
  and $T_{\mathrm{after}}=6~\MeV$, the dotted (blue in the web
  version) lines for
  $\rho_{\mathrm{X}}(10~\GeV)=8\,\rho_{\rad}(10~\GeV)$ and
  $T_{\mathrm{after}}=4.9~\MeV$.}
\label{Fig:EntropyProduction}
\end{center}
\end{figure}

In Fig.~\ref{Fig:EntropyProduction}
we show the evolution of $S$, $a^3\rho_{\mathrm{X}}$, and
$a^3\rho_{\rad}$ for two exemplary scenarios
respecting~(\ref{Eq:Tafter_limit}). The scale factor $a$ is normalized
by $a_{\mathrm{I}}\equiv a(10~\GeV)=1~\GeV^{-1}$ and the temperature
dependence of $g_*$ is taken into account as determined
in~\cite{Gondolo:1990dk}.
For $\rho_{\mathrm{X}}(10~\GeV)=0.1\,\rho_{\rad}(10~\GeV)$ and
$T_{\mathrm{after}}=6~\MeV$, $S$ increases by a factor of $\Delta=100$
as shown by the corresponding solid line.
For $\rho_{\mathrm{X}}(10~\GeV)=8\,\rho_{\rad}(10~\GeV)$ and
$T_{\mathrm{after}}=4.9~\MeV$, $S$ increases by a factor of
$\Delta=10^4$ as shown by the corresponding dotted (blue in the web
version) line.

We restrict our study to entropy production at late times,
$T_{\mathrm{before}}\simeq\TL \ll \TR$, so that the thermal production
of gravitinos is not affected. To work in a model independent way, we
assume that the production of gravitinos and NLSPs in the entropy
producing event, such as the direct production in decays of X, is
negligible.\footnote{The constraints discussed below shall therefore
  be considered as conservative bounds. For studies of gravitino
  production during an entropy producing event, we refer
  to~\cite{Kohri:2004qu} and references therein.}
Moreover, in this section, we focus on scenarios in which the
decoupling of the NLSP is not or at most marginally affected by
entropy production, i.e., either $\TR \gg T_{\mathrm{after}}\gg
T_{\freezeout}^{\NLSP}$ or $\rho_{\rad}\gg\rho_{\mathrm{X}}$ for
$T\gtrsim T_{\freezeout}^{\NLSP}$.
Thus, the thermally produced gravitino yield and---in the case of
entropy production after NLSP decoupling---also the non-thermally
produced gravitino yield are diluted:
\begin{equation}
  Y_{\gravitino}(T_{\mathrm{after}}) 
  =  
  \frac{S({\TL})}{S(T_{\mathrm{after}})}\,
  Y_{\gravitino}(\TL)
  \ .
\end{equation}

In the case of late-time entropy production \textit{before} the
decoupling of the NLSP, 
we parameterize this by writing
\begin{equation}
Y_{\gravitino}^{\TP}(T_0) 
= 
\frac{1}{\delta}\,
Y_{\gravitino}^{\TP}(\TL)
\ .
\end{equation}
In this case, $Y_{\NLSP}(T_0)$ and thereby
$\Omega_{\widetilde{G}}^{\NTP}$ and the BBN constraints remain
unaffected.

Conversely, in the case of late-time entropy production \textit{after}
the decoupling of the NLSP (and before BBN)
both, $Y_{\gravitino}^{\TP}(T_0)$ and $Y_{\NLSP}(T_0)$, are reduced:
\begin{eqnarray}
 Y_{\gravitino}^{\TP}(T_0) 
 &=&  
 \frac{1}{\Delta}\,Y_{\gravitino}^{\TP}(\TL)
 \nonumber \\
 Y_{\NLSP}(T_0) 
 &=& 
 \frac{1}{\Delta} Y_{\NLSP}(\TL)
\end{eqnarray}
Accordingly, $\Omega_{\widetilde{G}}^{\TP}$ and
$\Omega_{\widetilde{G}}^{\NTP}$ become smaller and the BBN constraints
can be relaxed. 

\begin{figure*}[t]
\begin{center}
\includegraphics[width=3.25in]{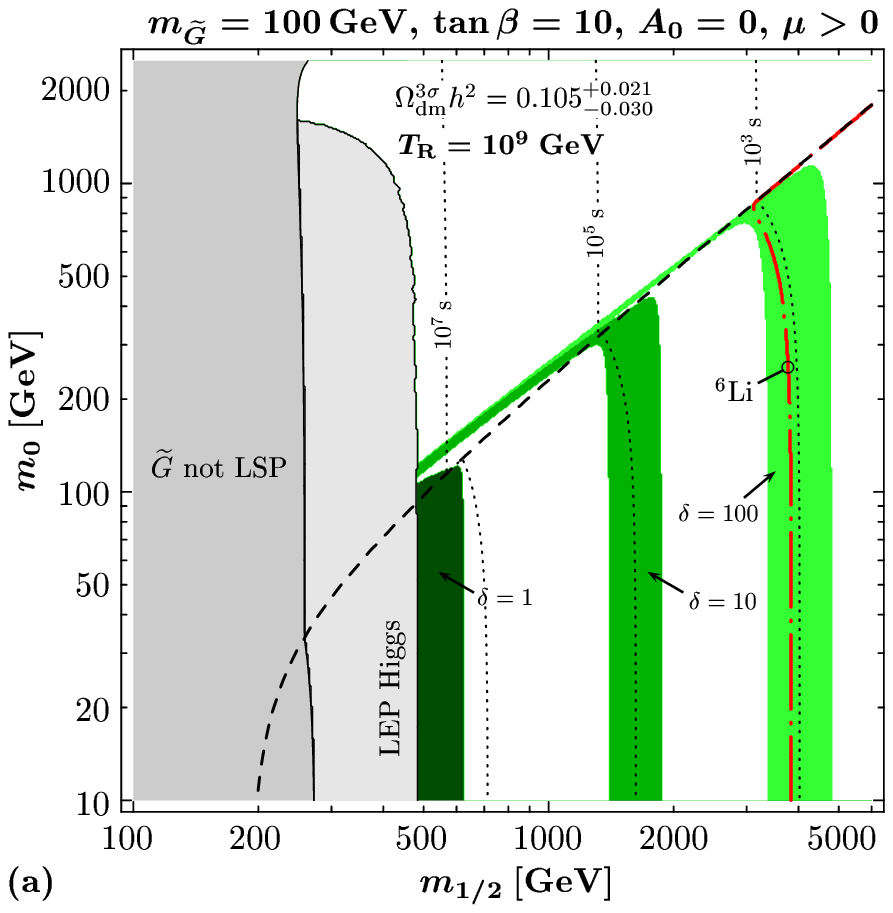} 
\hskip 0.75cm
\includegraphics[width=3.25in]{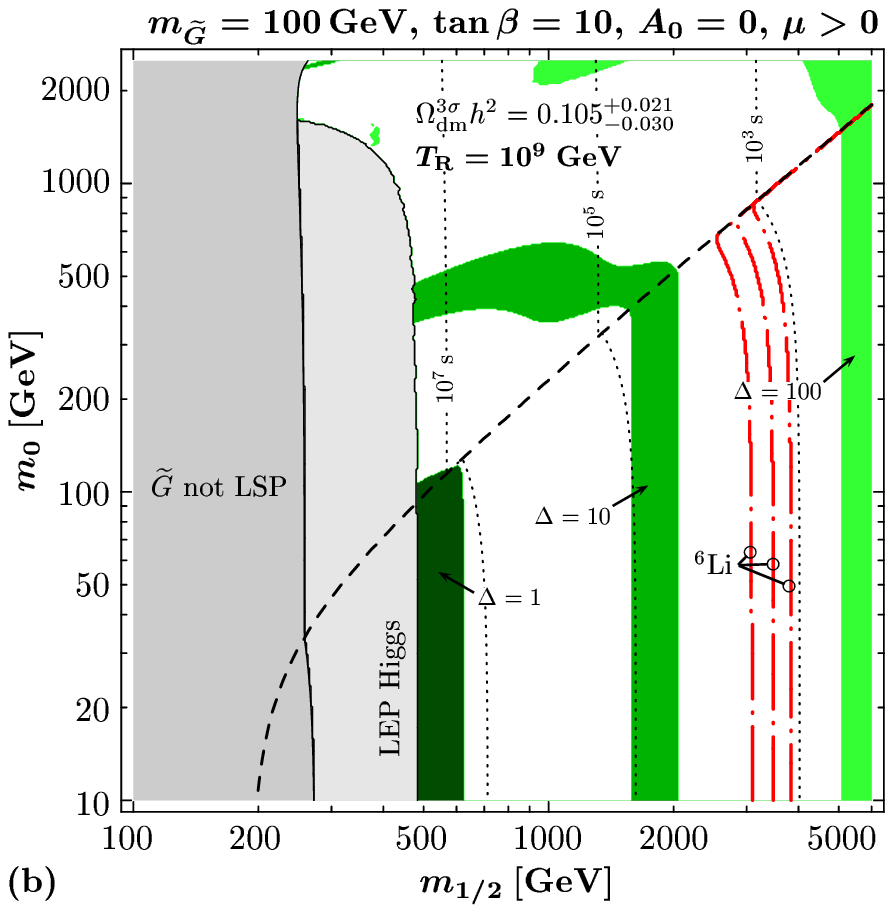} 
\vskip 0.75cm
\includegraphics[width=3.25in]{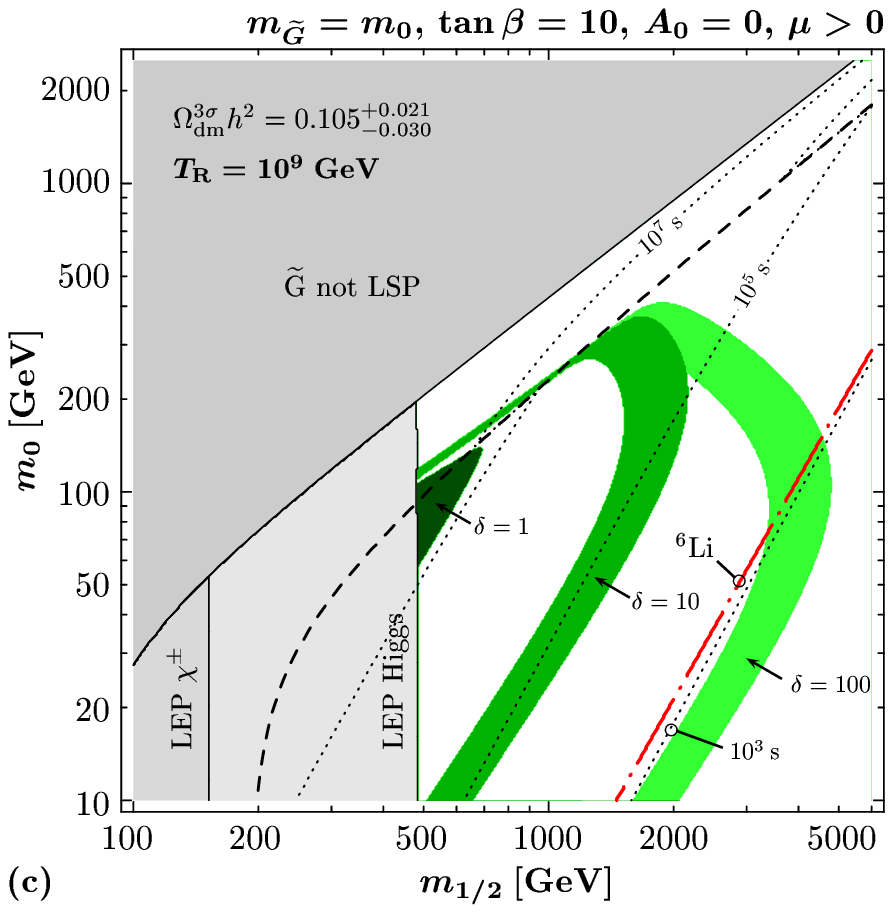} 
\hskip 0.75cm
\includegraphics[width=3.25in]{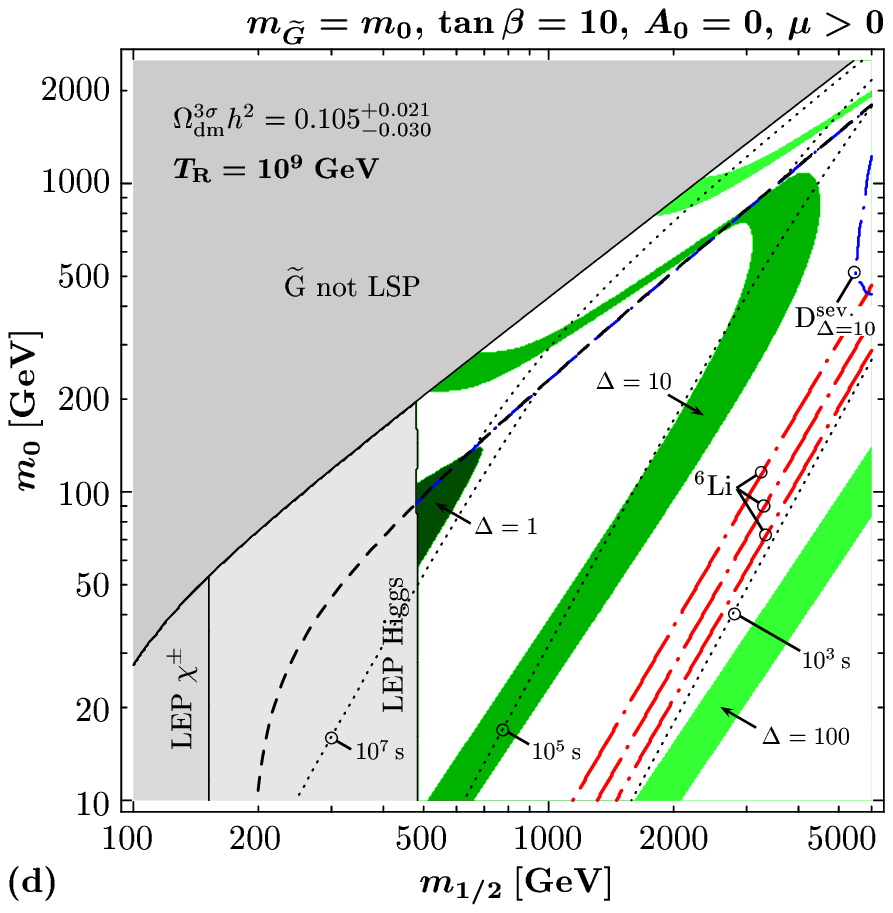} 
\caption{\small The effect of late-time entropy production before
  (left) and after (right) NLSP decoupling on regions in which
  $0.075\leq\Omega_{\gravitino}h^2\leq 0.126$ for $\TR=10^9~\GeV$.
  The $(m_{1/2},m_0)$ plane is shown for $\tan\beta=10$, $A_0=0$,
  $\mu>0$, $\mgr=100~\GeV$ (upper panels) and $\mgr=m_0$ (lower
  panels). The dark shaded (green in the web version) region is
  obtained without late-time entropy production $\delta=\Delta=1$. The
  medium and light shaded (green in the web version) bands are
  obtained with a dilution of $\Omega_{\gravitino}^{\TP}$
  ($\Omega_{\gravitino}^{\TP}+\Omega_{\gravitino}^{\NTP}$) by
  $\delta=10$ ($\Delta=10$) and $\delta=100$ ($\Delta=100$),
  respectively. The $\stau$ NLSP region to the right of the dot-dashed
  (red in the web version) line is cosmologically disfavored by the
  primordial $^6$Li abundance. Other curves and regions are identical
  to the ones in the corresponding panels of Fig.~\ref{Fig:CMSSMtB10}.
  The severe D constraint for $\Delta=10$ appears only in panel~(d).}
\label{Fig:CMSSMEntropyProductionI}
\end{center}
\end{figure*}

In Fig.~\ref{Fig:CMSSMEntropyProductionI}
we show how late-time entropy production before (left) and after
(right) NLSP decoupling affects the $^6$Li constraint and the region
in which $0.075\leq\Omega_{\gravitino}h^2\leq 0.126$ for
$\TR=10^9~\GeV$.  The $(m_{1/2},m_0)$ planes are considered for
$\tan\beta=10$, $A_0=0$, $\mu>0$, $\mgr=100~\GeV$ (upper panels) and
$\mgr=m_0$ (lower panels).  The dark shaded (green in the web version)
region is obtained without late time entropy production,
$\delta=\Delta=1$. The medium and light shaded (green in the web
version) bands are obtained with a dilution of
$\Omega_{\gravitino}^{\TP}$
($\Omega_{\gravitino}^{\TP}+\Omega_{\gravitino}^{\NTP}$) by
$\delta=10$ ($\Delta=10$) and $\delta=100$ ($\Delta=100$),
respectively. The dot-dashed (red in the web version) line illustrates
that the $^6$Li bound is independent of $\delta$, as shown in the
panels on the left-hand side, and becomes weaker (i.e., moves to the
left) with increasing $\Delta$, as shown in the panels on the
right-hand side. Other curves and regions are identical to the ones in
the corresponding panels of Fig.~\ref{Fig:CMSSMtB10}.
Note that we do not show the D constraint on late hadronic energy
injection since it is not sensitive to $\delta$ and vanishes already
for $\Delta=10$; an exception is the severe D constraint which still
appears for $\Delta=10$ in panel~(d).
BBN constraints on $\neutralino$ NLSP scenarios with entropy
production after NLSP decoupling will be studied elsewhere.

Comparing panels~(b) and~(d) of Fig.~\ref{Fig:CMSSMtB10} with
panels~(a) and~(c) in Fig.~\ref{Fig:CMSSMEntropyProductionI}, we find
that a dilution factor of $\delta=10$ ($100$) relaxes the $\TR$ bound
by a factor of 10 (100). Since the BBN constraints are unaffected by
$\delta$, the cosmologically disfavored range of NLSP masses cannot be
relaxed. With the dilution after NLSP decoupling, the relaxation of
the $\TR$ constraints is more pronounced.  Here also the
cosmologically disfavored range of NLSP masses can be
relaxed~\cite{Buchmuller:2006tt}. However, as can be seen in
panels~(b) and~(d) of Fig.~\ref{Fig:CMSSMEntropyProductionI}, the
$^6$Li bound is persistent. With a dilution factor of $\Delta=100$,
large regions of the $(m_{1/2},m_0)$ plane remain cosmologically
disfavored. For $\Delta\gtrsim 10^4$, however, the $^6$Li bound can be
evaded as will be shown explicitly below.

Figure~\ref{Fig:CMSSMEntropyProductionI} shows that inflation models
predicting, for example, $\TR=10^9~\GeV$ become allowed in the CMSSM
with gravitino dark matter for $\delta=\Delta\approx 100$. Here it is
not necessary to have late-time entropy production in the somewhat
narrow window between NLSP decoupling and BBN. This is different for
the viability of thermal leptogenesis in the considered scenarios
($T^{\gravitino}_{\freezeout}>\TR$) and for collider prospects as
discussed below.

%_____________________________________________________________________
\section{Thermal leptogenesis in the CMSSM with gravitino dark matter}
%_____________________________________________________________________

The constraint $\TR\lesssim 10^7~\GeV$ obtained in the considered CMSSM
scenarios for a standard cosmological history strongly disfavors
thermal leptogenesis. However, if entropy is released after NLSP
decoupling, a dilution factor of $\Delta\simeq 10^4$ can render
thermal leptogenesis viable for $\TR\simeq 10^{13}~\GeV$.

Standard thermal leptogenesis usually requires $\TR\gtrsim
10^9~\GeV$~\cite{Buchmuller:2004nz}.  However, late-time entropy
production dilutes the baryon asymmetry which is generated well before
NLSP decoupling,
\begin{equation}
\eta(T_{\mathrm{after}})
= 
\frac{1}{\Delta}\,
\eta(T_{\mathrm{before}})
\ .
\end{equation}
Therefore, the baryon asymmetry before entropy production must be
larger by a factor of $\Delta$ in order to compensate for the
dilution. For $\Delta\simeq 10^4$, this can be achieved in the case of
hierarchical neutrinos for
$M_{\mathrm{R}1}\sim\TR\simeq 10^{13}~\GeV$,
as can be seen in Fig.~7~(a) of Ref.~\cite{Buchmuller:2002rq} and in
Fig.~2 of Ref.~\cite{Buchmuller:2002jk}. Here $M_{\mathrm{R}1}$ is the
mass of the lightest among the heavy right-handed Majorana neutrinos.

In Fig.~\ref{Fig:EntropyProduction} the dotted (blue in the web
version) lines show a scenario in which a dilution factor of
$\Delta=10^4$ is generated in the out-of-equilibrium decay of a heavy
particle X.  Because of
$\rho_{\mathrm{X}}(10~\GeV)=8\,\rho_{\rad}(10~\GeV)$, the Hubble rate
can be enhanced already during the decoupling phase of the NLSP, which
leads to an increase of $T_{\freezeout}^{\NLSP}$ and
$Y_{\NLSP}(T_{\freezeout}^{\NLSP})$.
In the results shown below, we account for this by using a modified
version of the \texttt{micrOMEGAs} code.\footnote{The $Y_{\NLSP}$
  contours shown in Fig.~\ref{Fig:YNLSP} do not apply in this
  section.}
After entropy production, the net effect is still a significant
reduction of $Y_{\NLSP}(T_0)$.
For the same initial conditions, $\Delta=2\times 10^4$---and thereby
an additional reduction of $Y_{\NLSP}(T_0)$ by a factor of two---can
be achieved by lowering $T_{\mathrm{after}}$ from $4.9~\MeV$ down to
$2.5~\MeV$.

\begin{figure}[t]
\begin{center}
\includegraphics[width=3.25in]{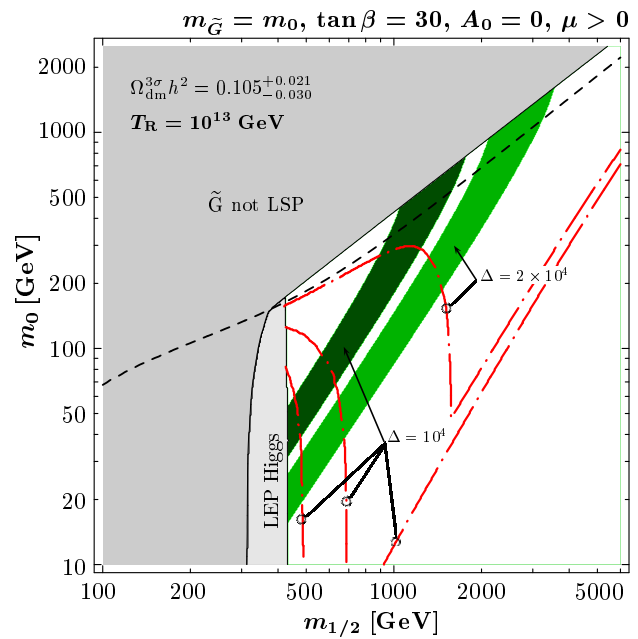} 
\caption{\small The effect of entropy production after NLSP decoupling
  for $\TR=10^{13}~\GeV$ and $\Delta\geq 10^4$ in the $(m_{1/2},m_0)$
  plane for $\tan\beta=30$, $A_0=0$, $\mu>0$, and $\mgr=m_0$.  The
  shaded (green in the web version) bands show the region in which
  $0.075\leq\Omega_{\gravitino}h^2\leq 0.126$ for $\Delta=10^4$ (dark)
  and $2\times 10^4$ (medium).  The dot-dashed (red in the web
  version) lines illustrate the corresponding evolution of the $^6$Li
  bound.  For $\Delta=10^4$, the regions below the associated two
  rightmost curves and to the right of the associated leftmost curve
  are allowed. For $\Delta=2\times 10^4$, the region below the line
  labeled accordingly is cosmologically allowed.}
\label{Fig:Leptogenesis}
\end{center}
\end{figure}

We consider these two scenarios 
for $\tan\beta=30$, $A_0=0$, $\mu>0$, and $\mgr=m_0$, in
Fig.~\ref{Fig:Leptogenesis}.
Here the shaded (green in the web version) bands indicate the region
in which $0.075\leq\Omega_{\gravitino}h^2\leq 0.126$ for
$\TR=10^{13}~\GeV$ and $\Delta=10^4$ (dark) and $2\times 10^4$
(medium). In addition, the corresponding evolution of the $^6$Li bound
is shown by the dot-dashed (red in the web version) lines. For
$\Delta=10^4$, the regions below the associated two rightmost curves
and to the right of the associated leftmost curve are allowed. For
$\Delta=2\times 10^4$, the cosmologically allowed region is the
$\stau$ NLSP region below the line labeled accordingly. The gray
regions are identical to the ones in Fig.~\ref{Fig:CMSSMtB30}.

We find that the $^6$Li bound cannot be evaded for the $\tan\beta=10$
scenarios even for $\Delta=2\times 10^4$ since $Y_{\NLSP}(T_0)$
becomes larger.  However, the $^6$Li bound given in Fig.~4 of
Ref.~\cite{Pospelov:2006sc} depends linearly\footnote{We thank
  M.~Pospelov for bringing this point to our attention.} on the
assumed limiting primordial abundance~(\ref{Eq:LiAbundance}) that is
subject to uncertainties; cf.\ Ref.~\cite{Jedamzik:2006xz}.
Accordingly, for a limiting abundance that is a factor of two above
the value given in~(\ref{Eq:LiAbundance}), one obtains the $^6$Li
bound labeled with $\Delta=2\times 10^4$ in
Fig.~\ref{Fig:Leptogenesis} for the scenario with $\tan\beta=30$ and
$\Delta=10^4$.

Scenarios with successful thermal leptogenesis in the $\stau$ NLSP
region are located preferably on the dark-shaded (dark green in the
web version) band and in the white corner to its left, in which even
slightly higher values of $\TR$ are possible for $\Delta=10^4$.  For
$\TR=10^{13}~\GeV$ and $\Delta\gg 10^4$, the generated baryon
asymmetry is diluted too strongly in order to explain the observed
baryon asymmetry.

As can be seen in Fig.~\ref{Fig:Leptogenesis}, the $\stau$ NLSP region
with
$500~\GeV\lesssim m_{1/2}\lesssim 700~\GeV$,
where $m_{\stau}\lesssim 200~\GeV$ (cf.~Fig.~\ref{Fig:YNLSP}), is no
longer disfavored by the $^6$Li bound provided $\Delta\gtrsim 10^4$.
Such scenarios are particularly promising since the long-lived $\stau$
NLSP could provide striking signatures of gravitino dark matter at
future
colliders~\cite{Buchmuller:2004rq,Brandenburg:2005he+X,Martyn:2006as,Hamaguchi:2006vu}.

%_____________________________________________________________________
\section{Conclusion}
%_____________________________________________________________________

Using the full gauge-invariant result for $\Omega_{\gravitino}^{\TP}$
to leading order in the Standard Model gauge
couplings~\cite{Pradler:2006qh}, we have studied bounds on
$T_{\Reheating}$ from the constraint $\Omega_{\gravitino}\leq
\Omega_{\CDM}$. Our results take into account the dependence of
$\Omega_{\gravitino}^{\TP}$ on the masses of the gauginos associated
with the Standard Model gauge group
SU(3)$_\Color\times$SU(2)$_\Weak\times$U(1)$_\Hypercharge$. This has
allowed us to explore the dependence of the $T_{\Reheating}$ bounds on
the gaugino-mass relation at the scale of grand unification
$M_{\GUT}$.

Within the CMSSM, we have explored gravitino dark matter scenarios and
the associated $\TR$ bounds for $\mgr\gtrsim 1~\GeV$ and for
temperatures as low as $10^7~\GeV$. Taking into account the
restrictive constraint from bound-state effects of long-lived
negatively charged staus on the primordial $^6$Li
abundance~\cite{Pospelov:2006sc}, we find that $\TR\lesssim 10^7~\GeV$
is the highest cosmologically viable temperature of the
radiation-dominated epoch in case of a standard thermal history of the
Universe.  This imposes a serious constraint on model building for
inflation.  Moreover, thermal leptogenesis seems to be strongly
disfavored in the considered regions of the CMSSM parameter space.

With late-time entropy release, the obtained limit $\TR\lesssim
10^7~\GeV$ can be relaxed.  For example, the dilution of the thermally
produced gravitino yield by a factor of $10$ relaxes the $\TR$ bound
by about one order of magnitude in regions where
$\Omega_{\gravitino}^{\TP}$ dominates $\Omega_{\gravitino}$.  In the
case of entropy production after NLSP decoupling, the yield of the
NLSP prior to its decay, $Y_{\NLSP}$, is reduced so that the BBN
constraints can be weakened. Although the $^6$Li bound is persistent,
we find that it disappears provided $Y_{\NLSP}$ is diluted by a factor
of $\Delta\gtrsim 10^4$.

We have discussed the viability of thermal leptogenesis in a
cosmological scenario with entropy production after NLSP decoupling.
We find that successful thermal leptogenesis can be revived in generic
regions of the CMSSM parameters space for 
$M_{\mathrm{R}1}\sim\TR \simeq 10^{13}~\GeV$ and $\Delta\gtrsim 10^4$,
where $M_{\mathrm{R}1}$ is the mass of the lightest among the heavy
right-handed Majorana neutrinos. 

Remarkably, for a dilution factor of $\Delta\gtrsim 10^4$, the $\stau$
NLSP region with $m_{\stau}\lesssim 200~\GeV$ reopens as a
cosmologically allowed region in the CMSSM with the gravitino LSP.  A
long-lived $\stau$ in this mass range could provide striking
signatures of gravitino dark matter at future
colliders~\cite{Buchmuller:2004rq,Brandenburg:2005he+X,Martyn:2006as,Hamaguchi:2006vu}.

\begin{acknowledgments}

\bigskip

We are grateful to S.~Blanchet, F.~Hahn-Woernle, M.~Pl\"umacher,
M.~Pospelov, G.~$\mbox{Raffelt}$, and Y.Y.Y.~Wong for valuable
discussions.
\end{acknowledgments}
%
% __________________________________________________________________

% __________________________________________________________________
%
%
\end{document}